\documentclass[conference]{IEEEtran}
\ifCLASSINFOpdf
\else
\fi
\usepackage[process=auto,cleanup={.dvi,.ps,.pdf,.log}]{pstool}
\begingroup
\makeatletter
\catcode`\#=11
\gdef\pstool@bitmap@opts{%
	-dAutoFilterColorImages#false
	-dAutoFilterGrayImages#false %
	-dColorImageFilter#/FlateEncode %
	-dGrayImageFilter#/FlateEncode 
}
\gdef\pstool@pspdf@opts{%
	-dPDFSETTINGS#/prepress %
	-dCompatibilityLevel#1.3 %
	-dEmbedAllFonts#true %
	-dSubsetFonts#true
}
\endgroup
\graphicspath{{fig/}}

\usepackage[caption=false, font=footnotesize,width=8cm,nearskip=5pt]{subfig}

\usepackage{amsfonts}
\usepackage{url}
\PassOptionsToPackage{hyphens}{url}
\usepackage{soul}

\usepackage[cmex10]{amsmath}
\usepackage{booktabs}
\newcommand{\PreserveBackslash}[1]{\let\temp=\\#1\let\\=\temp}

\usepackage{multicol}
\usepackage{multirow}
\usepackage{longtable}
\usepackage{tabularx}
\usepackage{graphicx}
\usepackage{tabulary}
\usepackage{amsfonts}
\usepackage{blkarray, bigstrut} %
\renewcommand{\vec}[1]{\ensuremath{\boldsymbol{#1}}} 

\makeatletter
\def\ps@IEEEtitlepagestyle{
	\def\@oddfoot{\mycopyrightnotice}
	\def\@evenfoot{}
}
\def\mycopyrightnotice{
	{\footnotesize
		\begin{minipage}{\textwidth}
			\textit{submitted to Power Systems Computation Conference (PSCC), 2022, Porto}
		\end{minipage}
	}
}

\begin{document}
%
\title{Multi-Horizon Planning of Multi-Energy Systems}



 \author{\IEEEauthorblockN{Tim Felling\IEEEauthorrefmark{1},
		Oliver Levers\IEEEauthorrefmark{1},
		Philipp Fortenbacher\IEEEauthorrefmark{1}}
	\IEEEauthorblockA{\IEEEauthorrefmark{1}Amprion GmbH (Transmission System Operator)\\
		Germany \\ \{tim.felling, oliver.levers, philipp.fortenbacher\}@amprion.net}
}

\maketitle

\begin{abstract}
In order to reach EU's goal of zero emissions in 2050, the energy system will go through a significant transition over the next decades. To substitute fossil energy carriers, renewable energy sources will be mainly integrated in the power system. Thereby, sector coupling will play a major role by making flexibility from other sectors such as heat or transport accessible to the power system. Planning the cost optimal transition requires a whole-system view over multiple horizons and across all sectors. This imposes the need for multi-energy system (MES) models coupled with multi-horizon investment models.  
This paper presents two multi-horizon planning approaches to determine the cost optimal pathway of a MES. As a major contribution, we propose a new method to incorporate technology-dependent learning cost curves in the planning problem and show that the resulting mixed-integer linear programming problem can be solved faster with a Benders decomposition technique as compared to a closed optimization. As a further contribution, we demonstrate the usefulness of our approach by showing the MES expansion pathway for a small German test system. 
\end{abstract}


\section{Introduction}

\subsection{Motivation}
Planning of infrastructure is rather a matter of decades than years. Thus, even though the target years 2045+ seem far away, it is of significant importance to start planning the energy transition to a carbon emission free system today. In addition, new required technologies still need time for technical development until they are ready for the market.\\
The key to zero emissions across all energy sectors in 2045+ lays in the reduction of the final energy consumption while increasing the share of electricity of the primary energy sources, as most of the renewable energy sources (RES), mainly photovoltaic (PV) and wind, provide electricity. 
As their infeed is highly fluctuating, an interlinked energy system will enable to better cope with flexibility and even store surplus energy over a longer period of time, e.g. by linking the power system to hydrogen, heat or gas system. In response, this helps to decarbonize these sectors. Hence, the individual sectors should no longer be considered on their own but rather together in an integrated manner. In consequence, expansions of energy infrastructures such as gas pipelines or power transmission lines need to be planned on an integrated level to outweigh their overall value across the interlinked sectors. 

In sum, this imposes the need for multi-energy system (MES) models that capture the interactions between several energy carriers and are able to plan the composition of the future energy system. Planning over the entire transition pathway requires special attention on the model choice~\cite{bell2019,MANCARELLA20141} as the value of infrastructures can only be assessed by models that run on high temporal and spatial resolutions. In addition, a yearly dispatch horizon is required to quantify the impacts on seasonal variations. Altogether, this requires to solve large-scale optimization problems. Therefore, it is crucial to bring their computational complexity to a manageable level by finding tractable formulations. 
\subsection{Related Work and Contribution}
Given the importance of the topic, several MES models with individual focal points have been developed in academia within the last years and found application in studies \cite{tennet,euco}. In contrast to macro-economic top-down models, bottom-up models use optimization to compute a least cost for the pathway of an energy system \cite{Pfluger2014}. Table \ref{mes_comparison} gives an overview over recent bottom-up models. To determine a certain transition pathway on which investment decisions are undertaken, either myopic approaches~\cite{PyPSA,fine} or closed approaches~\cite{times,primes} are applied. Closed approaches with perfect foresight have the advantage to obtain a cost optimal pathway, since they take the entire pathway into account, but induce higher complexity to the problem. Myopic approaches have the risk to lead to suboptimal results and stranded investments \cite{anyMOD,GERBAULET2019973,LOFFLER2019100422}. The decline of the installation cost per quantity as a function of the cumulative installed quantity is referred to as learning cost curves and is a realistic behavior of equipment producers. The modeling of this ability is even more complex.  
\begin{table}[t]
	\centering
	\caption{Comparison of existing multi-energy system models.}
	\label{mes_comparison}
	\resizebox{\columnwidth}{!}{\begin{tabular}{l|l|l|l|l|l|l|l}
		Model & Dispatch  & \multicolumn{3}{c|}{Investment} & \multicolumn{2}{c|}{Resolution} &  Problem\\
			  & Horizon   & Model  & Horizon	& Learning  & temporal & spatial & Class \\
		\toprule
		PyPSA \cite{PyPSA} & yearly & myopic & multiple & - & high & high & LP \\
		\hline
		Enertile \cite{enertile} & yearly & closed & multiple & - & high & high & LP\\
		\hline
		FINE \cite{fine} & weekly & myopic & multiple & - & high & high & MILP \\
		\hline
		Dimension \cite{dimension} & weekly & closed & single & - & high & high & LP \\
     	\hline
		anyMOD \cite{anyMOD} & yearly & closed & multiple & - & high & high & LP \\
		\hline
		TIMES \cite{times} & yearly & closed & multiple & x & low & low & LP,MILP \\
		\hline
		PRIMES \cite{primes}& yearly & closed & multiple & x & low & low & NLP\\
		\hline
		LISA & yearly & closed & single,   & x & high & high & LP, \\  
		    &   & & multiple & & & & MILP  \\
		\bottomrule
	\end{tabular}}
\end{table}

The contribution of this paper is two-fold. First, we present a tractable MES model capturing seasonal MES operation and incorporating a single-horizon investment model. We call our model LISA, which stands for light integrated energy system analysis. Second, we extend the LISA model by two multi-horizon planning approaches to determine the optimal investment decisions over a given transition path. The pathway consists of several horizons each starting with the supporting year, in which the investment decision takes place. In contrast to existing myopic approaches \cite{PyPSA,victoria2020}, we solve the optimization problem (OP) with a closed approach and, thus, take future investment decisions over the course of the entire transition path into account. Our planning approaches consider end-of-horizon effects and value discounting. 
The first approach considers linear cost functions associated with the investment decisions and is formulated as a linear programming (LP) problem. As a major contribution, our second approach is able to incorporate technology-dependent learning cost curves closing the gap between fine ~\cite{PyPSA,enertile,fine,dimension} and coarse~\cite{times,primes} granular energy system models. Their consideration results in concave cost functions, which makes the planning problem non-convex. We propose a method that translates the non-convex problem into a mixed-integer linear programming (MILP) problem by using piecewise-linear cost curves. Both approaches are either solved in a closed optimization (CO) or by a Benders decomposition (BD) method. 

The remainder of this paper structures as follows. Section~\ref{sec:Method} lays out the methodology, in which we present the formulation of our MES model LISA. Section~\ref{sec:Application} presents the application case of the five-node system whose results are shown in Section~\ref{sec:Results}. Section~\ref{sec:Conclusion} gives a conclusion and an outlook on future work. 
 


\section{Method}\label{sec:Method}
\subsection{Multi-Energy System Dispatch Problem} \label{subsec:MP_Disp}
Since we aim to simulate the seasonal operation of large MES, we formulate a multi-period dispatch problem by using tractable formulations from \cite{geidl2007}. In addition, we model the MES components in an unified framework using the concepts of \cite{Heussen2012}. 
\subsubsection{Sector Coupling}
First, we define a set of energy carriers $\alpha,\beta,\gamma,\delta,\omega \in \mathcal{S}$ = \{electricity, methane, hydrogen, heat, external\}, in which the MES can interact. We introduce generation and demand power vectors $\vec{p}^{\mathrm{g}}_j \in \mathbb{R}^{n^\mathrm{g}_j \times 1}, \vec{p}^{\mathrm{d}}_j \in \mathbb{R}^{n^\mathrm{d}_j \times 1}$ that are associated with the energy system components for each sector $j \in \mathcal{S}$. We also define $(\vec{p}^{\mathrm{g}}_j, \vec{p}^{\mathrm{d}}_j) \in \vec{p}_j$. The coupling between the sectors follows the same way as suggested in \cite{geidl2007} by introducing sector converters $\vec{c}_{ij}$ with $(i,j) \in \mathcal{S}$. The vector $\vec{c}_{ij}$ represents the efficiency for the energy conversion. The coupling can be mathematically expressed by  
\begin{equation}
	\left[ \begin{array}{cccc} \vec{I} & -\vec{c}_{\alpha\beta} & \hdots &  -\vec{c}_{\alpha\omega} \\
								-\vec{c}_{\beta\alpha} & \vec{I} & \ddots &  -\vec{c}_{\beta\omega}  \\
								 \vdots & \ddots & \ddots & \vdots  \\
							 	-\vec{c}_{\omega\alpha} & -\vec{c}_{\omega\beta} & \hdots &  \vec{I} \end{array} \right]
		\left[ \begin{array}{c} \vec{p}_\alpha \\ \vec{p}_\beta \\ \vdots \\ \vec{p}_\omega    \end{array} \right] = \vec{0} \ , \label{eq:sectormodel}					 	 
\end{equation}
where $\vec{I}$ is the identity matrix.

\subsubsection{Multi-Sectoral Net Transfer Capacity (NTC) Transport Model}
To model the sectoral energy distribution over a network, we map the sectoral nodal injections on a transport flow model by defining 
\begin{equation}
\vec{C}^{\mathrm{g}}_j\vec{p}^{\mathrm{g}}_j - \vec{C}^{\mathrm{d}}_j\vec{p}^{\mathrm{d}}_j	= \vec{C}^{\mathrm{ft}}_j\vec{p}^{\mathrm{t}}_j \quad,
\label{eq:transportmodel}
\end{equation}
where $\vec{p}^{\mathrm{t}}_j \in \mathbb{R}^{n^\mathrm{l}_j \times 1}$ are the sectoral transport flows linked with the incidence matrix $\vec{C}^{\mathrm{ft}}_j \in \mathbb{Z}^{n^\mathrm{b}_j \times n^\mathrm{l}_j}$. It defines the mapping of $n^\mathrm{l}_j$ from-to links to their corresponding nodal injections. The incidence matrices $\vec{C}^{\mathrm{g}}_j \in \mathbb{Z}^{n^\mathrm{b}_j \times n^\mathrm{g}_j}, \vec{C}^{\mathrm{d}}_j \in \mathbb{Z}^{n^\mathrm{b}_j \times n^\mathrm{d}_j}$ project $n^\mathrm{g}_j$ generation variables and $n^\mathrm{d}_j$ demand variables to $n^\mathrm{b}_j$ nodal injections.

\subsubsection{Description of Energy System Component Models}
We use the unified modeling approach from \cite{Heussen2012} to describe any component of the energy system. We adopt the generic description from \cite{Heussen2012} and expand it to more sectors in our notation by  
\begin{equation}
	\vec{e}_j(k+1) = \vec{e}_j(k) -\vec{H}_j^{\mathrm{g}}\vec{p}^{\mathrm{g}}_j(k) \\ + \vec{H}_j^{\mathrm{d}}\vec{p}^{\mathrm{d}}_j(k) + \vec{\xi}_j(k) \label{eq:stormodel}
\end{equation}
to configure the system behavior of a component in the sector $j \in\mathcal{S}$. The vector $\vec{e}_j(k)$ represents the energy levels of storage devices. The matrices $\vec{H}_j^{\mathrm{g}}= \mathrm{diag}\{\vec{\eta}^\mathrm{g}_j\}^{-1}, \vec{H}_j^{\mathrm{d}}= \mathrm{diag}\{\vec{\eta}^\mathrm{d}_j\}$ incorporate the charging and discharging efficiencies $\vec{\eta}^\mathrm{d}_j,\vec{\eta}^\mathrm{g}_j$ and the vector $\vec{\xi}_j(k)$ is an external variable to exogeniously describe inflows or outflows. As presented in \cite{Heussen2012} this generic formulation allows us to parameterize any demand, storage and generation component by the extension to exclusively consider components from other sectors such as hydrogen or methane. In particular, we can model the aggregated behavior of components that provide flexibility such as heat pumps, electrolyzers, electric vehicles, demand side applications, thermal and battery storage devices. It is also possible to configure curtailable and intermittent RES such as PV plants, wind turbines etc. 

\subsubsection{Multi-Period Dispatch Problem Formulation}
The interactions between the energy system components can be formulated by an intersectoral multi-period dispatch problem. The corresponding OP determines the optimal sectoral setpoints $\vec{p}_j(k) \in \vec{P}_j$, sectoral transport flows $\vec{p}^\mathrm{t}_j(k) \in \vec{P}^\mathrm{t}_j$ and storage trajectories $\vec{e}_j(k) \in \vec{E}_j$  over a given dispatch horizon $N$ at time steps $k$ with $(N,k) \in \mathcal{T}$ to minimize the operating cost. The problem is
\begin{equation}
	\begin{array}{lll}
		J^{*} = & \multicolumn{2}{l} {\displaystyle\min_{(\vec{E}_j,\vec{P}_j,\vec{P}^\mathrm{t}_j) \in \vec{X}} \ \left(\sum\limits_{k \in \mathcal{T}} \sum\limits_{j \in \mathcal{S}} \ \vec{c}_j^T \vec{p}_j(k) \right)}   \\
		& \multicolumn{2}{l} {\text{s.t.} \ \forall k\in \mathcal{T}, \forall j\in \mathcal{S}} \\
		& \text{(a)} & \eqref{eq:sectormodel},\eqref{eq:transportmodel},\eqref{eq:stormodel}  \\
		& \text{(b)} &\sum\limits_{k \in \mathcal{T}}\vec{a}_\mathrm{CO2}^T \vec{p}_j^{\mathrm{g}}(k) \leq b_\mathrm{CO2}  \\
		& \text{(c)} & \vec{C}^{\mathrm{g}}_\gamma p_{\gamma}^{\mathrm{g}}(k) \leq \vec{C}^{\mathrm{g}}_\beta p_{\beta}^{\mathrm{g}}(k) b_\mathrm{h2} \\
		& \text{(d)} & \vec{e}_j(N) = \vec{e}_j(1) \\ 
		& \text{(e)} & \vec{e}_j(1) = \vec{e}_j^\mathrm{set} \\
		& \text{(f)} & \vec{0} \leq \vec{e}_j(k) \leq \overline{\vec{e}}_j \\
		& \text{(g)} & \underline{\vec{\xi}}_j(k)\leq \vec{p}_j(k) \leq \overline{\vec{\xi}}_j(k) \\
		& \text{(h)} &\vec{0} \leq \vec{p}^\mathrm{t}_j(k) \leq \overline{\vec{b}}^{\mathrm{ntc}}_j  \quad, 
	\end{array} \label{eq:dispatchproblem}
\end{equation}
where $\vec{c}_j$ defines the fuel cost for the primary energy sources and $\vec{a}_\mathrm{CO2}$ specifies their corresponding green house gas (GHG) emission factors. The constraint (\ref{eq:dispatchproblem}b) ensures that GHG emissions is below a chosen GHG threshold $b_\mathrm{CO2}$. We also capture hydrogen blending into the methane system by limiting the infeed of produced hydrogen to a certain fraction $b_\mathrm{h2}$ of the instantaneous methane infeed. This is accomplished by applying (\ref{eq:dispatchproblem}c) where the incidences $\vec{C}^{\mathrm{g}}_{\gamma,\beta}$ map the associated component outputs to the nodal injections. Constraints (\ref{eq:dispatchproblem}d-e) ensure that the initial and terminal storage levels are equal and set to a predefined initial value. Constraints (\ref{eq:dispatchproblem}f-h) bound the storage levels and transport flows to their operating ranges and (\ref{eq:dispatchproblem}g) incorporates inflows, outflows, profile series for RES and loads, and power ratings of the energy system components.

For the sake of compactness \eqref{eq:dispatchproblem} can be formulated as a standard LP problem as
\begin{equation}
	\begin{array}{lll}
		J^{*} = & \multicolumn{2}{l} {\displaystyle\min_{\vec{X}} \ \vec{c}^T\vec{X} }   \\
		& \multicolumn{2}{l} {\text{s.t.}} \\
		& \vec{A}\vec{X} \leq \vec{b}  \\
		& \underline{\vec{X}} \leq \vec{X} \leq  \overline{\vec{X}} \quad, 
	\end{array}
\end{equation}
\noindent where the vector $\vec{c}$ captures the cost objective of \eqref{eq:dispatchproblem}, the matrix $\vec{A}$ in combination with vector $\vec{b}$ represent the inequalities and equalities of (\ref{eq:dispatchproblem}a-d) and $\underline{\vec{X}},\overline{\vec{X}}$ incorporate the lower and upper bounds of (\ref{eq:dispatchproblem}e-h).    

\subsection{Equivalent annual cost problem}\label{subsec:SingleHorizon}
Given the presented multi-period dispatch problem, we now extend the OP by introducing new investment variables $\vec{z} \in \mathbb{R}^{n_z \times 1}$ that eventually extend the upper bounds of selected components (4g-h) as presented in \eqref{eq:singe_invest_prob}:
\begin{equation}\label{eq:singe_invest_prob}
	\vec{A}\vec{X}+\vec{G}\vec{z}\le \vec{b} \quad.
\end{equation}
Therein, the new matrix  $\vec{G}$ contains one vector for each investment variable. For regular capacities the hourly value constitutes in (-1), while variables with profiles are considered with their profile values. Optionally, the investment can be bounded by a maximal additional capacity $ \overline{z}_{j}$:
\begin{equation}
	0 \leq  z_j  \leq  \overline{z}_{j} \ .
\end{equation}
Finally, considering the annual additional costs $c_\mathrm{inv}$ for $z$ yields in objective function of the equivalent annual cost problem (EACP):
\begin{equation}\label{eq:MinEACP}
	\begin{array}{lll}
		J^{*} = & \multicolumn{2}{l} {\displaystyle\min_{\vec{X,z}} \ \vec{c}^T\vec{X} + \vec{c}_{\mathrm{inv}}^{T}\vec{z} }   \\
		& \multicolumn{2}{l} {\text{s.t.}} \\
		& \vec{A}\vec{X} +\vec{G}\vec{z} \leq \vec{b}  \\
		& \underline{\vec{X}} \leq \vec{X} \leq  \overline{\vec{X}} \\
		& \vec{0} \leq \vec{z} \leq \overline{\vec{z}} \quad. 
	\end{array}
\end{equation}
\begin{figure}[t]
	\centering
	\def\svgwidth{\columnwidth}
	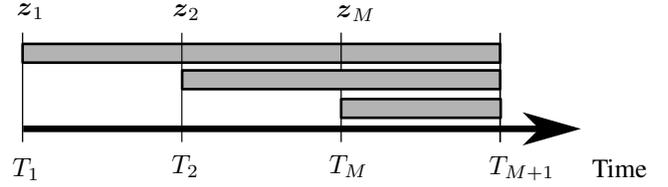
	\caption{Multi-horizon planning over the course of the entire transition path. The pathway consists of $M$ horizons each starting with the supporting year $T$, in which the investment decision $\vec{z}$ takes place. The gray boxes illustrate the coupling of the investment decisions with the matrix $\tilde{\vec{G}}$ over the expansion path.}
	\label{fig:multi-horizon}
	\vspace{-3mm}
\end{figure}
\subsection{Multi-Horizon Investment and Dispatch Problem}\label{subsec:MultipleHoizon}
Having introduced the formulation of the EACP for a single year, the following subsections present the extension to a multi-horizon OP. 

To this end, we define a pathway $\mathrm{T}=\{ T_1,...,T_M \}$ with $M=\#\mathrm{T}$ investment horizons. Notably, one $T_m$ can represent either a single year or serve as an average year for the horizon between the years $T_m$ and $T_{m+1}$. Thus,  $\mathrm{T}$  does not have to comprise consecutive years. This results in $\tilde{\vec{X}} = \left[  \vec{X}_1  \dots  \vec{X}_M \right]^T$ and $\tilde{\vec{z}} = \left[ \vec{z}_1 \dots \vec{z}_M \right]^T$ investment decisions and constitutes in the multi-horizon OP given in \eqref{eq:multistageInvestment}: \\
\begin{equation}
	\begin{array}{lll}
		J^{*} = & \multicolumn{2}{l} {\displaystyle\min_{\vec{\tilde{X}},\vec{\tilde{z}}} \ \sum_{m=1}^{M}w_m^\mathrm{disp}\vec{c}_m^T\vec{X}_m +f_m\left(\vec{z}_m \right)}   \\
		& \multicolumn{2}{l} {\text{s.t.}} \\
		& \left[\begin{array}{ccccc}
		   \tilde{\vec{A}} & \tilde{\vec{G}} \\
			\vec{0} & \vec{F}\\ 		     
		\end{array}\right]
		\left[\begin{array}{ccccc}
			\tilde{\vec{X}} \\
			\tilde{\vec{z}}\\ 		     
		\end{array}\right]\le \left[\begin{array}{ccccc}
			\tilde{\vec{b}} \\
			\vec{b}_\mathrm{F}\\ 		     
		\end{array}\right]\\\\
		& \underline{\tilde{\vec{X}}} \leq \tilde{\vec{X}} \leq  \overline{\tilde{\vec{X}}} \\ 
		& \vec{0} \leq \tilde{\vec{z}} \leq  \overline{\tilde{\vec{z}}} \quad, 
	\end{array}\label{eq:multistageInvestment}
\end{equation}
where $\tilde{\vec{A}}$ comprises the single investment horizons $\vec{A}_1 ... \vec{A}_M$ as block matrices on the diagonal. The previously introduced $\vec{G}$ is extended by one column for each investment horizon $m$ and, thus, yields in $\tilde{\vec{G}}$. This matrix allows us to couple investments from previous horizons with subsequent ones, which is shown in Fig.~\ref{fig:multi-horizon}. Notably, each investment decision is a variable over the entire path in contrast to myopic approaches, where earlier investment decisions become constants for subsequent horizons.
The matrix $\vec{F}$ optionally defines a pathway of bounds for total investment capacities over the entire horizon by coupling the individual $z_{j,m}$.

Moreover, as we now consider both investment and dispatch costs over multiple horizons, in contrast to the EACP, we discount and weigh the costs for the single horizons. To this end, the dispatch costs $\vec{c}_m$ of each investment horizon $T_m$ have to be multiplied by the discounting factor $w_m^\mathrm{disp}$: 
\begin{equation}
	\begin{array}{ll}
		w_m^\mathrm{disp}&=\sum_{t=T_m}^{T_{m+1}-1}\left( \frac{1}{1+w} \right)^{t-T_0}\\
		&=\left(\frac{1}{1+w}\right)^{T_m-T_0}\cdot\frac{1-\left(\frac{1}{1+w}\right)^{T_{m+1}-T_m}}{w} \ ,
	\end{array}
\end{equation}
where $w$ is the weighted average cost of capital (WACC) factor. Due to discounting and the limited lifetime $T_\mathrm{L}$ of investments the investment costs are multiplied by
\begin{equation}
	w_m^\mathrm{inv}=\frac{1}{(1+w)^{T_m-T_0}}\cdot\frac{T_{\mathrm{M+1}}-T_m}{T_{\mathrm{L}}} \ .
\end{equation} 
The cost functions $f_m(\vec{z}_m)$ represent either linear investment costs, as previously introduced in \eqref{eq:MinEACP}, or specific costs with learning effects. Those are introduced subsequently.  

\subsubsection{Convex Cost Functions}
In a first approach we consider constant specific costs $\vec{c}_{\mathrm{inv},m}$ for each investment horizon. Consequently, we get linear cost functions $f_m(\vec{z}_m)=w_m^\mathrm{inv}\vec{c}_{\mathrm{inv},m}^T\vec{z}_m$ and our multi-horizon investment problem remains a standard LP problem.

\subsubsection{Non-convex Cost Functions}
\begin{figure}[t]
	\centering
	\includegraphics[width = 8.5cm] {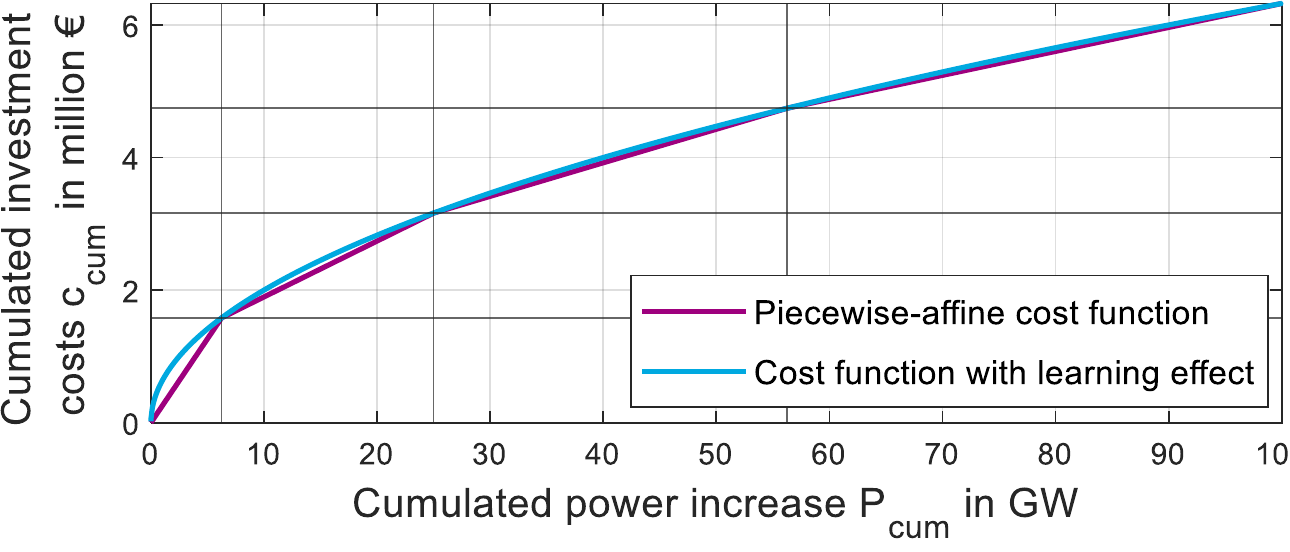}
	\caption{Modeling of learning curves with piecewise-affine cost functions.}
	\label{fig:nonconvex_function}
	\vspace{-2mm}
\end{figure}

In our second approach we assume that some specific costs are subject to learning effects. This allows us to model the costs more realistically. Especially for new technologies, prices decrease depending on the capacities produced. This dependence of specific costs on capacities leads to non-convex cost functions.
To this end, we define $c_{\mathrm{inv},0}$ to be the initially specific costs and $P_0$ to be the initially installed capacity of a random technology. Subsequently, we can express the learning effect depending on an increase in investment capacity  $P$ as an exponential regression based on \cite{barreto2001technological} with learning index $r$. We assume that the learning effects occur for the aggregated installed capacity across all considered regions:
\begin{equation}
	c_\mathrm{inv}(P)= \dfrac{c_{\mathrm{inv},0}}{P_0^{-r}}\cdot\left(P_0+P \right)^{-r} .
\end{equation}
 By integrating the specific costs we get the cumulated costs required for the multi-horizon investment problem:
\begin{equation}
	c(P)=\dfrac{c_{\mathrm{inv},0}\cdot P_0}{1-r}\cdot\left( \left(1+\dfrac{P}{P_0}\right)^{1-r}-1\right).
\end{equation}
Because of the convex and monotonically decreasing specific costs the accumulated costs are concave and monotonically increasing. To incorporate these nonlinear cost functions, we use piecewise-affine functions, which are visualized in Fig. \ref{fig:nonconvex_function}. The consideration of these concave functions turns the former LP problem with linear cost terms to a MILP problem. In our approach we use piecewise-affine functions with equidistant $N_\mathrm{pw}$ set points $y_n$ and special ordered sets of type~2 (SOS2) to integrate the functions in the MILP problem. SOS2 are ordered sets of variables, where at maximum two variables could be nonzero. In addition, these variables have to be adjacent. The cost functions can be formulated as follows based on \cite{williams2013model}:
\begin{equation}
	\begin{array}{lll}
		c_\mathrm{pw}(P) =& \multicolumn{2}{l}{ \sum_{n=1}^{N_\mathrm{pw}}\gamma_{n}\cdot c(y_n)} \\
		& \multicolumn{2}{l} {\text{s.t.}} \\ 
		& \text{(a)} &P= \vec{\gamma}^T\vec{y}\\	
		& \text{(b)} &\vec{\gamma}^T \vec{1}  =  1\\	
		& \text{(c)} & \vec{\gamma}\le\left[\begin{array}{ccccc}
					1 & 0 & \cdots & 0 \\
					1 & 1 & \ddots & \vdots \\
					0 & \ddots & \ddots & 0 \\
					\vdots & \ddots & 1 & 1 \\     
					0 &\cdots & 0 & 1 \\
				\end{array}\right]\vec{\delta}\\	
					& \text{(d)} &\vec{\delta}^T\vec{1} = 1\\
		& \text{(e)} &\vec{\gamma}\in [0,1]^{N_\mathrm{pw}}\\
		& \text{(f)} &\vec{\delta}\in \{0,1\}^{N_\mathrm{pw}-1} \ .
	\end{array}\label{eq:nonconvexcosts}
\end{equation}
Therein, $\vec{\gamma}$ is the SOS2 and $\vec{\delta}$ is an integer vector. So far we considered cumulated investment costs. In a multi-horizon investment problem it is necessary to get the investment costs for each horizon. The added capacities $\Delta P_m = P_m - P_{m-1}$  are considered and the investment costs in this horizon are formulated as $c_{\mathrm{pw},m}(\Delta P_m) = 	w_m^\mathrm{inv} \cdot \left( c_\mathrm{pw}(P_m)-  c_\mathrm{pw}(P_{m-1})\right)$. To calculate the total investment costs we link \eqref{eq:nonconvexcosts} for each horizon and get 
\begin{equation}
c_{\mathrm{pw}} = \sum_{m=1}^{M}\left(w_{m}^\mathrm{inv}-w_{m+1}^\mathrm{inv}\right)\cdot\sum_{n=1}^{N_\mathrm{pw}}\gamma_{n,m}\cdot c(y_n)
\end{equation}
as an additional term of the objective function and $\Delta P_m=\vec{\gamma}_m^T\vec{y}-\vec{\gamma}_{m-1}^T\vec{y}$ as a new constraint of the OP. Additionally the constraints (\ref{eq:nonconvexcosts}b-f)  apply for all $m\in \{1,...,M\}$ and are incorporated additionally to the problem. While the resulting OP can be solved by a CO, the structure of the problem is suitable for the application of a BD. The corresponding formulation of the BD is presented subsequently. 


\subsubsection{Benders decomposition}\label{sec:BD}
The BD is an iterative method where the OP is decomposed in one master problem and $M$ subproblems based on \cite{conejo2006decomposition}. The subproblems can be formulated as
\begin{equation}
	\begin{array}{lll}
		J^{*} = &  {\displaystyle\min_{\vec{X}_m} w_m^{\mathrm{disp}}\vec{c}_m^T\vec{X}_m}   \\
		& {\text{s.t.}} \\
		& \left[\begin{array}{ccccc}
			\vec{A}_m & \vec{G}_m \\
			\vec{A}_{\mathrm{inv},m} & \vec{G}_{\mathrm{inv},m}\\ 		     
		\end{array}\right]
		\left[\begin{array}{ccccc}
			\vec{X}_m \\
			\tilde{\vec{z}}\\ 		     
		\end{array}\right]\le \left[\begin{array}{ccccc}
			\vec{b}_m \\
			\vec{b}_{\mathrm{inv},m}\\ 		     
		\end{array}\right]\\\\
		& \underline{\vec{X}}_m \leq \vec{X}_m \leq  \overline{\vec{X}}_m \\ 
		& \tilde{\vec{z}}=\tilde{\vec{z}}^*:\vec{\lambda}, 		
	\end{array}
\end{equation}
where $\vec{z}^*$ is the solution of the master problem
\begin{equation}
	\begin{array}{lll}
		J^{*} = & \displaystyle\min_{\vec{z}_1,...,\vec{z}_M,\alpha} \sum_{m=1}^{M} f_m\left(\vec{z}_m,  \vec{\delta}_m, \vec{\gamma}_m \right) + \alpha  \\
		&  {\text{s.t.}} \\
		& 	\vec{F}	\ 
			\left[\vec{z} \
			\vec{\delta} \
			\vec{\gamma} \right]^T\le \vec{b}\\
		& \vec{0} \leq \alpha \leq  \infty \\ 
		& \vec{0} \leq \vec{z} \leq  \overline{\vec{z}}\\
		& \vec{\gamma}\in [0,1]^{N_\mathrm{pw}}\\
		& \vec{\delta}\in \{0,1\}^{N_\mathrm{pw}-1}, 
	\end{array}
\end{equation}
which is extended after each iteration $l$ by the Benders cut
\begin{equation}
	 \sum_{m=1}^{M}-\vec{\lambda}_{m,l}^*\tilde{\vec{z}}_m\le \sum_{m=1}^{M}-\vec{\lambda}_{m,l}^*\tilde{\vec{z}}_{m,l}^*-w_m^{\mathrm{disp}}\vec{c}_{m}^T\vec{X}^*_{m,l},
\end{equation}
where $\vec{X}^*_{m,l} $ is the solution of subproblem $m$ in iteration $l$ and $\vec{\lambda}^*_{m,l}$ is the Lagrange multiplier of the equality $\tilde{\vec{z}}_m = \tilde{\vec{z}}_m^*$ in iteration $l$. As soon as the difference between the upper bound $\sum_{m=1}^{M}\vec{c}_{m}^T\vec{X}^*_{m,l}$ and the lower bound $\alpha_l^*$ is below a given tolerance $\varepsilon$ the BD is stopped.
Consequently, the decomposition yields a MILP problem as a master problem for non-convex cost functions and LP problems as subproblems for both cost functions.

\section{Application}\label{sec:Application}
\subsection{Multi-Energy Test System}
To compare the CO and BD solution approaches, we create a four-region MES model of Germany supplemented by a fifth region that represents the surrounding European countries. It is possible to im- and export electricity from this region to a fixed wholesale price that eventually, depending on the investment horizon, causes carbon emissions. Moreover, the model consists of an power transmission grid among the five regions and a methane grid between the four German regions. Note that the purpose of the test system is to demonstrate the validity of the results and the transferability for larger systems. Due to the simplifications, the test system rather provides an indicative transition pathway that could be different to a realistic one. The considered period of the MES starts 2030 and ends in 2060 with corresponding investment decisions in 2030, 2040 and 2050. In line with European climate goals,  the energy system has to be carbon neutral in 2050. The pathway of the $CO_2$ budget is given in Table \ref{tbl:co2-budget}. Further parameters of the test MES are summarized in Table \ref{tbl:parameter}. 
\begin{table}[t]
	\centering
	\caption{Exogenous parameters of the multi-energy test system.}
	\begin{tabular}{c||c|c|c}
		& 2030& 2040 & 2050\\ \hline
		$CO_2$-Budget per year  [Million tons]	& 200	& 90	& 0 	\\
		Power $CO_2$-Emission factor (EU) [t/MWh] 	& 0.4	& 0.2	& 0 	\\
		Wholesale electricity price (EU) [€/MWh]	& 67	& 83	& 87 	
	\end{tabular}
	\label{tbl:co2-budget}
\end{table}

\begin{table}[t]
	\centering
	\caption{Configuration of the multi-horizon investment model.}
	\begin{tabular}{c||c}		
		parameter & value \\
		\hline
		MES transition pathway   	& 2030-2060				\\
		investment years			& 2030, 2040, 2050		\\ \hline
		WACC						& 7 $\%$ 				\\ \hline
		dispatch horizon	  	    & 8760 h				\\ \hline
		dispatch time resolution    & 1h \\
		set points of pw-affine fct.& 100	\\ \hline
		\textbf{Initial capacity} [GW]\\
		Electrolysis & 0\\
		PV& 91.3\\
		Wind Offshore& 17\\
		Wind Onshore 	& 81.5\\		
	\end{tabular}
	\label{tbl:parameter}
\end{table}

Regarding the investment variables, we allow the endogenous investments in electrolyzers, PV, wind onshore and offshore capacities
as well as in NTC of the power transmission grid. Furthermore, it is possible to invest in NTC of a newly constructed German hydrogen grid. Due to the complexity of non-convex cost functions, we assume that CAPEX of onshore turbines and NTC can be described by linear cost functions. 

%
For electrolyzers, PV and wind offshore assets we model learning curves by piecewise-affine functions. The CAPEX for the linear cost functions and the learning indices are given in Table \ref{tbl:endo_lin2}.




\begin{table}[t]
	\centering
	\caption{CAPEX and learning index of possible investments.}
	\begin{tabular}{c||c|c|c||c|c}
			& \multicolumn{3}{c||}{CAPEX [€/kW]}&life time&learning\\	
		 & 2030& 2040 & 2050 &$T_L$&   index $r$ [\%]\\
		\hline
		Wind Offshore		& 2691	& 2326	& 2011	& 20	& 31.9	\\
		PV					& 900	& 793	& 700	& 25	& 20	\\ 
		Electrolysis		& 651	& 479	& 353  	& 15	& 13.3	\\
		Wind Onshore		& 1200	& 1112	& 1031	& 20	& -		\\
		\hline		
			& \multicolumn{3}{c||}{CAPEX [€/(kW 100 km)]}&life time&learning\\	
		& 2030& 2040 & 205 0&$T_L$ &   index $r$ [\%]\\
		\hline 
		Power Grid			& 40	& 40	& 40  	& 40	& -		\\
		Hydrogen Grid		& 40	& 40	& 40    & 40	& -
	\end{tabular}
	\label{tbl:endo_lin2}
\end{table}

\subsection{Complexity Analysis}
We use Gurobi 9.0 to solve the OPs. To analyze the complexity of our approaches we vary the horizon length $N$ in the dispatch problem from 20 to 8760 simulated hours. We analyze the runtime behaviour for the convex and nonconvex cost function case using the BD and CO solution strategy.

\section{Results}\label{sec:Results}
Having introduced the application case, the following two sections present the complexity analysis of the two solution approaches and the actual results of the multi-horizon MES. 

\subsection{Complexity Analysis}\label{subsec:ResultsComplex}

\begin{figure}[t]
	\centering
	\includegraphics[width = 8.5cm] {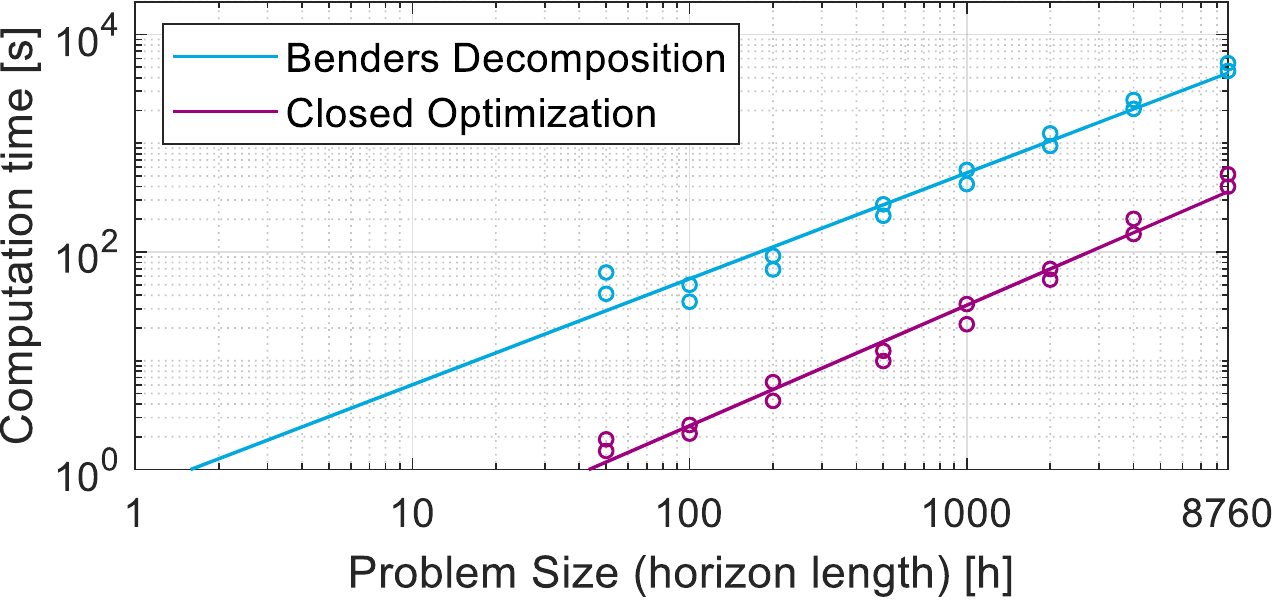}
	\caption{Computation time for the convex cost function case solved by LP as a function of the problem size. The closed optimization (CO) strategy performs better than the Benders decomposition (BD) yielding lower computation times.}
	\label{fig:convex_runtime}
\end{figure}

Figure \ref{fig:convex_runtime} shows the log-log regressions of the computation time of the convex cost function case (LP problem) depending on the horizon length for both the CO and BD. It indicates that the CO method is faster than the BD. Due to their different slopes it can be assumed that an intersection could occur for larger problem sizes. However, this intersection point could not be determined given the calculation time.

\begin{figure}[t]
	\centering
	\includegraphics[width = 8.5cm] {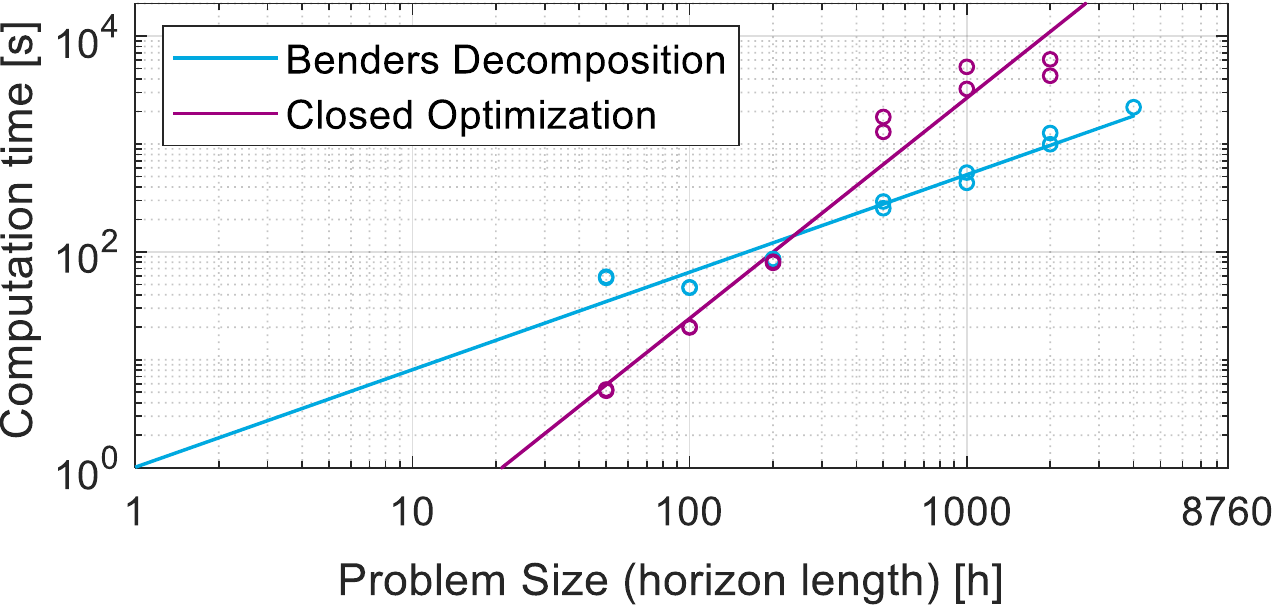}
	\caption{Computation time for the non-convex cost function case solved by MILP as a function of the problem size. For larger problem sizes the Benders decomposition outperforms the closed optimization strategy, which is indicated by the intersection to lower computation times.}
	\label{fig:nonconvex_runtime}
\end{figure}

In contrast, Fig. \ref{fig:nonconvex_runtime} shows that the two complexity curves for non-convex cost functions intersect approximately at an horizon length of 220h. Thus, in that case the BD is advantageous for larger horizon lengths. As shown in Section \ref{sec:BD}, the BD divides the problem in relatively large LP subproblems and one small MILP problem. Consequently, the complexity of BD only slightly increase in comparison to the LP problem and, thus, can be solved relatively quick. In turn, when using the CO method we gain a very large MILP problem which is hard to solve in an acceptable time frame. 

\subsection{Resulting MES Expansion Pathway }
\begin{figure}[t]
	\centering
	\includegraphics[width = 8.5cm] {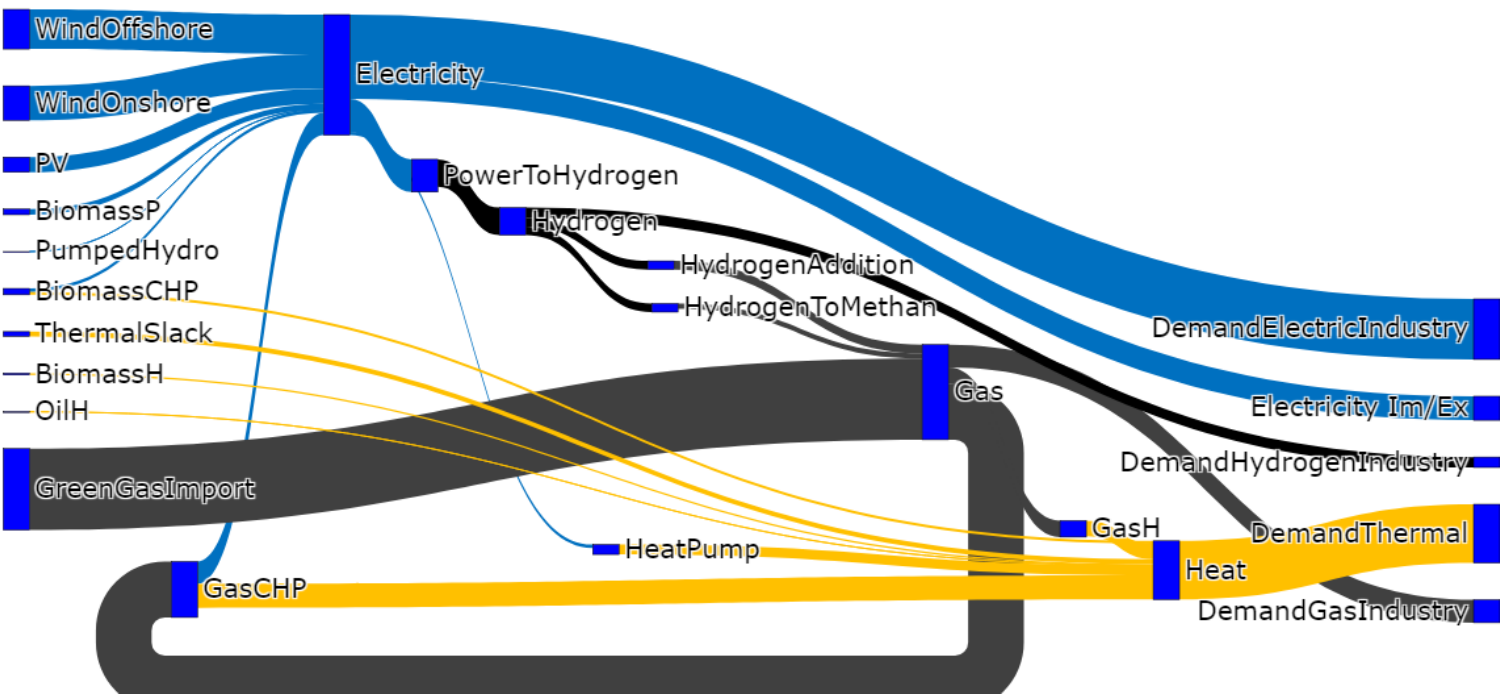}
	\caption{Energy flow diagram example of yearly MES dispatch results in 2050.}
	\label{fig:Sankey}
\end{figure}

After comparing the complexities of the BD and CO, we now compare the results of both approaches and demonstrate the impact of learning curves for MES. 
Because of the inapplicability of the CO for large horizons we now compare two set-ups: First, we evaluate the overall functionality of the CO and BD by comparing their results for the OP without learning curves. Second, we investigate the results of the BD with and without learning curves. The key results of the in total three model runs are shown in Table \ref{tbl:results}. An exemplary yearly dispatch that visualizes the interplay of the four sectors is given in the sankey diagram in Fig. \ref{fig:Sankey}. 

\begin{table*}[t!]
	\centering
	\caption{Results of the MES expansion pathway for the investment horizons from 2030 to 2050.}
	\begin{tabular}{c|c|c||c|c|c|c||c|c|c|c||c|c|c|c}
		\multicolumn{3}{c||}{}&\multicolumn{4}{c||}{MES with learning curves (BD)} & \multicolumn{4}{c||}{MES with linear costs (BD)} & \multicolumn{4}{c}{MES with linear costs (CO)} \\
		\hline
		\multicolumn{2}{c}{}	&unit	&2030&2040&2050&Total&	2030&2040&2050&Total&	2030&2040&2050&Total\\
		\hline		
		\hline
		\multicolumn{2}{c|}{\textbf{Investments}}&&&&&&&&&&&&&\\
		Electrolysis	&Total					& GW	& 28.0	& 38.3	& 28.7	& 95.0	& 3.4	& 43.3	& 42.1	& 88.7	& 3.4	& 43.8	& 42.1	& 89.3	\\
		&North					& GW	& 17.6 	& 27.4 	& 0.6	& 45.5	& 0.7	& 26.8	& 20.0	& 47.5	& 0.8	& 25.8	& 17.1	& 43.7	\\
		&East					& GW	& 2.6	& 6.5	& 11.3	& 20.3	& 0.8	& 6.4	& 7.0	& 14.2	& 0.7	& 5.4	& 9.0	& 15.1	\\ 
		&West					& GW	& 7.8	& 2.1	& 15.7 	& 25.6	& 1.4	& 7.4	& 8.5 	& 17.2	& 1.3	& 9.7	& 10.8 	& 21.8	\\
		&South					& GW	& 0		& 2.3	& 1.2 	& 3.6	& 0.5	& 2.7	& 6.6 	& 9.8	& 0.6	& 2.9	& 5.3 	& 8.7 	\\		
		\multicolumn{2}{l|}{PV} 				& GW	& 0		& 0		& 40.9	& 40.9	& 0		& 25.4	& 50	& 75.4	& 0		& 34.8	& 50	& 84.8	\\
		\multicolumn{2}{l|}{Wind Onshore} 		& GW 	& 25.9	& 40	& 40 	& 105.9	& 0		& 40	& 40 	& 80	& 0		& 40	& 40 	& 80	\\
		\multicolumn{2}{l|}{Wind Offshore}		& GW 	& 24	& 24	& 24 	& 72	& 24	& 24	& 24  	& 72	& 24	& 24	& 24 	& 72	\\
		\multicolumn{2}{l|}{Power Grid}			& GW 	& 1.1	& 5.5	& 6.3 	& 13.0	& 0.1	& 3.1	& 20.6	& 23.8 	& 1.5	& 1.9	& 7.1	& 10.6	\\
		\multicolumn{2}{l|}{Hydrogen Grid}		& GW 	& 0.6	& 0		& 8.2 	& 8.8	& 0.1	& 0	 	& 19.8 	& 19.9	& 0.1	& 0.1	& 0.8	& 1.0	\\
		
		\hline
		\hline
		\multicolumn{2}{c|}{\textbf{Costs}}		& 	 	&		&		&		&		&		&		&		&		&		&		&		&		\\	
		\multicolumn{2}{c|}{Operating costs}	& bn € 	& 654.5 & 304.1	& 249.7	& 1208.3& 691.6	& 316.6	& 251.6	& 1259.8& 691.4	& 313.8	& 250.5	& 1255.7\\
		\multicolumn{2}{c|}{Investment costs}	& bn € 	& 128.4	& 59.0	& 21.2 	&  208.6& 79.5	& 78.5	& 22.9	& 180.9	& 79.7	& 82.1	& 22.8	& 184.6	\\
		\hline
		\multicolumn{2}{c|}{Total costs}		& bn € 	& 782.9	& 363.1	& 270.9	& \textbf{1416.9}	& 771.2	& 395.0	& 274.5	& \textbf{1440.7}	& 771.1 & 395.9  & 273.3  & \textbf{1440.3}\\
	\end{tabular}
	\label{tbl:results}
\end{table*}

\subsubsection{Closed Optimization and Benders Decomposition}\label{subsec:ResultsCOvsBD}
Comparing the results of the CO and the BD we would expect similar results and in any case the same objective values. In fact, the two objective values of the MES with linear cost functions, which represent the system costs from 2030 to 2060, differ by 0.4 million € which is only a relative difference of 0.02\% and within the set accuracy of the BD.

\begin{figure}[t]
	\centering
	\vspace{-0.4cm}
	\includegraphics[width = 8.5cm] {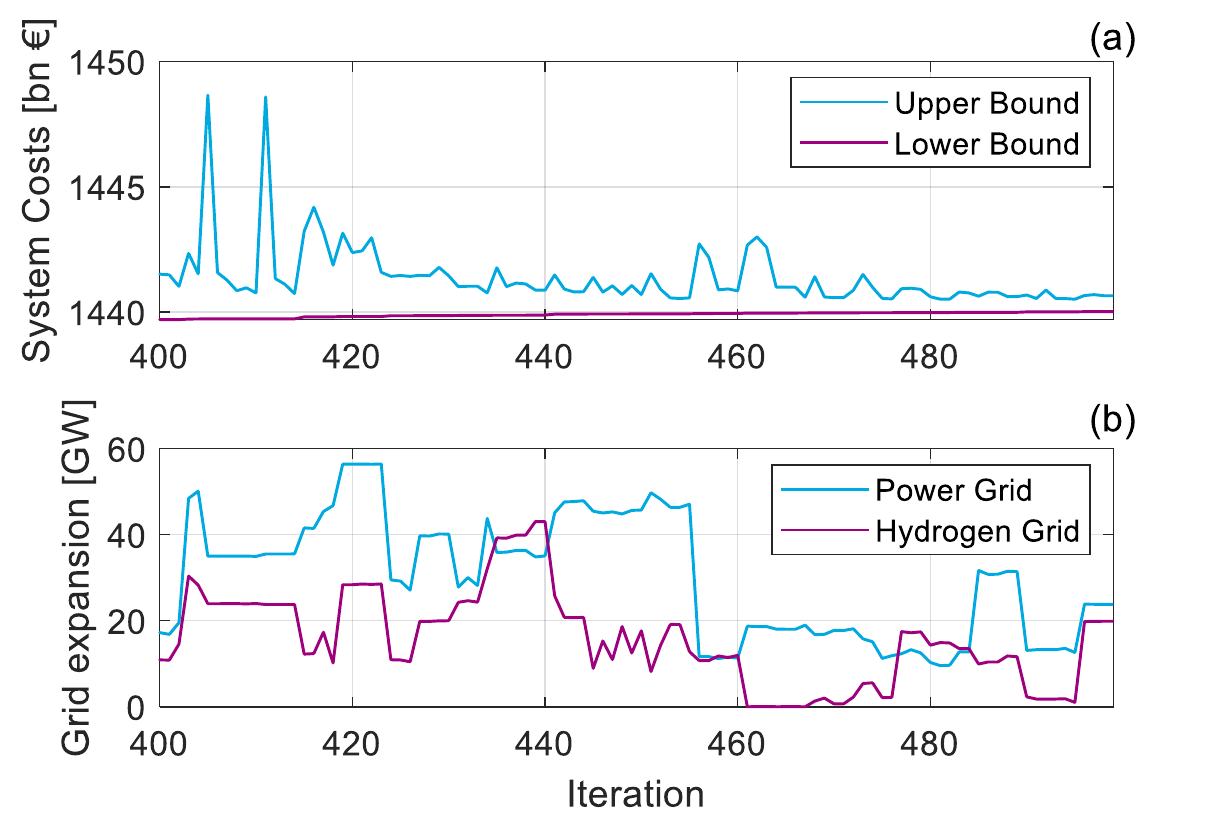}
	\caption{Grid expansion over last iterations of Benders decomposition.}
	\label{fig:BendersConvergence}
	\vspace{-0.5cm}
\end{figure}

Nevertheless, although the objective values are almost identical, the investment paths are quite distinct. Especially the constructed hydrogen grids differ. When solving the OP with BD, the optimal solution results in 19.9 GW of newly constructed hydrogen NTC. 
In contrast, the solution of the CO results in only 1 GW. Moreover, the CO identifies higher investments in photovoltaics. This leads to a more decentralized power generation and consequently in less need for transmission capacities. 
One reason for the different pathways with almost similar costs is delivered by Fig. \ref{fig:BendersConvergence} (a), which shows the investments in the hydrogen and power grid for different number of iterations, while Fig. \ref{fig:BendersConvergence} (b) presents the according objective value (system costs). Even though the upper and lower bounds of the BD converge to the same final objective value, the characteristics of network expansion differ significantly even within the last iterations. This suggests that there might be quite different solutions, flat optima, for very similar objective values.
 

\subsubsection{Linear Cost Functions and Non-convex Cost Functions} \label{subsec:ResultsMILP}
In contrast to the previous comparison, a difference in both objective value and transition path between the linear and non-convex cost functions would be quite expected. 
In fact, the objective value of the MES with learning effects is by 23.8 billion € lower in comparison to the objective value of to the MES with linear cost functions. That constitutes in a relative difference of 1.7 \%. Hence, the objectives are less similar as in the comparison before but still close to each other. Nevertheless the investment paths are quite different. Particularly the installed capacities of PV and onshore wind turbines differ widely. 

When using learning curves, the learning effect of PV seems to be too low to trigger higher and earlier investments in PV. Thus, the optimizer rather chooses to invest at an earlier horizon in wind onshore capacities in contrast to the MES with linear cost curves.
The most innovative technology we consider is electrolysis which constitutes in the highest impact of learning effects. This affects the investment decisions: While the total installed capacity of the two investment paths is quite similar, the annually values differ. Given the learning effects, the optimizer chooses to invest already in 2030 in electrolysis what  reduces the CAPEX later on. Comparing the regionalization of the investments, the installed power of electrolysis in northern Germany is higher in comparison to the MES with linear cost functions in 2030 and 2040. In contrast, investments in western Germany are much lower during this period. This is closely interlinked and driven by the concentration of renewable energies, e.g. in the northern areas of Germany due to the above mentioned investments in wind onshore energy in 2030.


\section{Conclusion and Outlook}\label{sec:Conclusion}
In this paper we present LISA, a tractable MES model capable of multi-horizon planning with both linear and non-convex cost functions. In addition, we present two solution strategies, a BD and a CO, to solve the multi-horizon investment problems. 

In terms of a complexity analysis, we show that each solution approach has its advantage. The CO is advantageous for linear cost functions while the BD is more suited for the non-convex cost variant. In the case of linear cost functions, both solution strategies arrive at objective values that are within the range of accuracy. Both transition paths are plausible, yet there are major differences in the (grid) investment decisions. This implies the need for further investigation of these flat optima and of the robustness of solutions. In case of different cost functions, we compare the corresponding transition paths determined by the BD. While the big picture of both transition paths is very similar, we identify an impact of learning effects. Finally, future work could incorporate a fixed $\mathrm{CO}_2$ budget for the whole transition in contrast to fixed thresholds for every investment horizon and a decommissioning path for the energy system assets.


\bibliographystyle{IEEEtran}
\bibliography{literature}

\begin{thebibliography}{10}
\providecommand{\url}[1]{#1}
\csname url@samestyle\endcsname
\providecommand{\newblock}{\relax}
\providecommand{\bibinfo}[2]{#2}
\providecommand{\BIBentrySTDinterwordspacing}{\spaceskip=0pt\relax}
\providecommand{\BIBentryALTinterwordstretchfactor}{4}
\providecommand{\BIBentryALTinterwordspacing}{\spaceskip=\fontdimen2\font plus
\BIBentryALTinterwordstretchfactor\fontdimen3\font minus
  \fontdimen4\font\relax}
\providecommand{\BIBforeignlanguage}[2]{{%
\expandafter\ifx\csname l@#1\endcsname\relax
\typeout{** WARNING: IEEEtran.bst: No hyphenation pattern has been}%
\typeout{** loaded for the language `#1'. Using the pattern for}%
\typeout{** the default language instead.}%
\else
\language=\csname l@#1\endcsname
\fi
#2}}
\providecommand{\BIBdecl}{\relax}
\BIBdecl

\bibitem{bell2019}
G.~Hawker and K.~Bell, ``\BIBforeignlanguage{English}{Making energy system
  models useful: good practice in the modelling of multiple vectors},'' vol.~9,
  no.~1, Jan. 2020.

\bibitem{MANCARELLA20141}
P.~Mancarella, ``{MES} (multi-energy systems): An overview of concepts and
  evaluation models,'' \emph{Energy}, vol.~65, pp. 1--17, 2014.

\bibitem{tennet}
\BIBentryALTinterwordspacing
{Tennet, Gasunie, FZ Juelich, IAEW Aachen, DBI}, ``Phase ii - pathways to
  2050,'' Tech. Rep., 2020. [Online]. Available:
  \url{https://www.tennet.eu/fileadmin/user_upload/Company/Publications/Technical_Publications/200204_Phase_II_Project_report.pdf}
\BIBentrySTDinterwordspacing

\bibitem{euco}
\BIBentryALTinterwordspacing
{European Comission}, ``A clean planet for all,'' Tech. Rep., 2018. [Online].
  Available:
  \url{https://eur-lex.europa.eu/legal-content/EN/TXT/?uri=CELEX:52018DC0773}
\BIBentrySTDinterwordspacing

\bibitem{Pfluger2014}
B.~Pfluger, ``Assessment of least-cost pathways for decarbonising europe's
  power supply : a model-based long-term scenario analysis accounting for the
  characteristics of renewable energies,'' Ph.D. dissertation, Karlsruhe
  Institute of Technology (KIT), May 2014.

\bibitem{PyPSA}
\BIBentryALTinterwordspacing
T.~Brown, J.~H\"orsch, and D.~Schlachtberger, ``{PyPSA: Python for Power System
  Analysis},'' \emph{Journal of Open Research Software}, vol.~6, no.~4, 2018.
  [Online]. Available: \url{https://doi.org/10.5334/jors.188}
\BIBentrySTDinterwordspacing

\bibitem{fine}
L.~Welder, D.~Ryberg, L.~Kotzur, T.~Grube, M.~Robinius, and D.~Stolten,
  ``Spatio-temporal optimization of a future energy system for
  power-to-hydrogen applications in germany,'' \emph{Energy}, vol. 158, pp.
  1130--1149, 2018.

\bibitem{times}
\BIBentryALTinterwordspacing
R.~Loulou, ``Documentation for the times model part i,'' International Energy
  Agency (IEA), Tech. Rep., 2016. [Online]. Available:
  \url{https://iea-etsap.org/docs/Documentation_for_the_TIMES_Model-Part-I_July-2016.pdf}
\BIBentrySTDinterwordspacing

\bibitem{primes}
\BIBentryALTinterwordspacing
{Energy-Economy Environment Modelling Laboratory}, ``Primes model version
  2018,'' NTU Athens, Tech. Rep., 2018. [Online]. Available:
  \url{https://e3modelling.com/wp-content/uploads/2018/10/The-PRIMES-MODEL-2018.pdf}
\BIBentrySTDinterwordspacing

\bibitem{anyMOD}
L.~Göke, ``A graph-based formulation for modeling macro-energy systems,''
  \emph{Applied Energy}, vol. 301, p. 117377, Nov 2021.

\bibitem{GERBAULET2019973}
C.~Gerbaulet, C.~{von Hirschhausen}, C.~Kemfert, C.~Lorenz, and P.-Y. Oei,
  ``European electricity sector decarbonization under different levels of
  foresight,'' \emph{Renewable Energy}, vol. 141, pp. 973--987, 2019.

\bibitem{LOFFLER2019100422}
K.~Löffler, T.~Burandt, K.~Hainsch, and P.-Y. Oei, ``Modeling the low-carbon
  transition of the european energy system - a quantitative assessment of the
  stranded assets problem,'' \emph{Energy Strategy Reviews}, vol.~26, p.
  100422, 2019.

\bibitem{enertile}
C.~Bernath, G.~Deac, and F.~Sensfuß, ``Impact of sector coupling on the market
  value of renewable energies – a model-based scenario analysis,''
  \emph{Applied Energy}, vol. 281, p. 115985, 2021.

\bibitem{dimension}
J.~Richter, ``Dimension - a dispatch and investment model for european
  electricity markets,'' Energiewirtschaftliches Institut an der Universitaet
  zu Koeln (EWI), EWI Working Papers 2011-3, 2011.

\bibitem{victoria2020}
M.~Victoria, K.~Zhu, T.~Brown, G.~B. Andresen, and M.~Greiner, ``Early
  decarbonisation of the european energy system pays off,'' \emph{Nature
  Communications}, vol.~11, no. 6223, pp. 1--9, 2020.

\bibitem{geidl2007}
M.~Geidl, G.~Koeppel, P.~Favre-Perrod, B.~Klockl, G.~Andersson, and
  K.~Fröhlich, ``Energy hubs for the future,'' \emph{IEEE Power and Energy
  Magazine}, vol.~5, no.~1, pp. 24--30, 2007.

\bibitem{Heussen2012}
K.~Heussen, S.~Koch, A.~Ulbig, and G.~Andersson, ``Unified system-level
  modeling of intermittent renewable energy sources and energy storage for
  power system operation,'' \emph{IEEE Systems Journal}, vol.~6, no.~1, pp.
  140--151, 2012.

\bibitem{barreto2001technological}
T.~L. Barreto~G{\'o}mez, ``Technological learning in energy optimisation models
  and deployment of emerging technologies,'' Ph.D. dissertation, ETH Zurich,
  2001.

\bibitem{williams2013model}
H.~P. Williams, \emph{Model building in mathematical programming}.\hskip 1em
  plus 0.5em minus 0.4em\relax John Wiley \& Sons, 2013.

\bibitem{conejo2006decomposition}
A.~J. Conejo, E.~Castillo, R.~Minguez, and R.~Garcia-Bertrand,
  \emph{Decomposition techniques in mathematical programming: engineering and
  science applications}.\hskip 1em plus 0.5em minus 0.4em\relax Springer
  Science \& Business Media, 2006.

\end{thebibliography}

\end{document}